\documentclass[aps,prc,superscriptaddress,groupedaddress,showpacs]
{revtex4}
\usepackage{graphicx}

\newcommand{\bra}[1]{\mbox{$\langle #1|$}}
\newcommand{\ket}[1]{\mbox{$|#1\rangle$}}
\newcommand{\braket}[2]{\mbox{$\langle #1 | #2 \rangle$}}
\begin{document}
FNT/T-2003/08, MKPH-T-03-10

\title{NN final-state interaction in two-nucleon knockout from  $^{16}O$}
 \author{M.\ Schwamb}
\address{Dipartimento di Fisica Nucleare e Teorica
dell'Universit\`a degli Studi di  Pavia, I-27100 Pavia, Italy 
and Institut f\"ur Kernphysik,
Johannes Gutenberg-Universit\"at, D-55099 Mainz, Germany}
\author{ S.\ Boffi, C.\ Giusti and  F.\ D.\  Pacati}
\address{Dipartimento di Fisica Nucleare e Teorica
dell'Universit\`a degli Studi di
  Pavia and Istituto Nazionale di Fisica Nucleare,
Sezione di Pavia, I-27100 Pavia, Italy}

\begin{abstract}
\noindent
 The influence  of the mutual interaction between the two outgoing
 nucleons (NN-FSI) in electro-  and photoinduced two-nucleon knockout 
 from $^{16}O$ has been  investigated perturbatively. It turns out that
 the effect of NN-FSI depends on the kinematics
 and on the type of reaction considered. The effect is generally larger in pp-
 than in pn-knockout and in electron induced than in photoinduced reactions.
  In superparallel kinematics NN-FSI leads   in the $(e,e'pp)$ channel to a
 strong increase of the cross section, that is mainly due to a strong 
 enhancement of the $\Delta$-current contribution. In pn-emission, however, 
 this effect is partially cancelled by a destructive 
 interference with the seagull current. For photoreactions NN-FSI is 
 considerably reduced in superparallel kinematics and can be
 practically  negligible in specific  kinematics.\\

\end{abstract}

\maketitle

\section{Introduction}
\label{intro}

 The independent particle shell model (IPM), which describes a nucleus as a  
 system  of nucleons moving in a mean field,  is able to reproduce  the 
 basic features of nuclear structure if effective NN-interactions 
 are employed, but using realistic interactions it fails to describe
the binding energy of nuclei.
 This failure is a consequence of the strong short-range 
 components of the interaction, which are necessary  to reproduce NN 
data and which
 induce into the nuclear wave function correlations beyond the mean field
 description. Thus, a careful evaluation of the short-range correlations (SRC) 
 is needed to describe nuclear properties in terms of a realistic 
NN-interaction
 and provide profound insight into the structure of the hadronic interaction 
 in the nuclear medium \cite{MuP00}.
 
 A  powerful tool for the investigation of SRC is the electromagnetically 
 induced  two-nucleon knockout since the probability that a real 
 or a virtual photon is absorbed by a pair of nucleons should be a direct 
 measure for the correlations between these nucleons\cite{Gott,Bof96}. 
 This simple picture, however, has to be modified because  additional
 complications have to be taken into account, such as competing mechanisms, 
 like  contributions of two-body currents as well as the final state 
 interaction (FSI) between the two outgoing nucleons and the residual nucleus. 

 As a logical consequence of the
 complexity of the problem, a combined study of all four possible reactions
$(e,e'pp)$, $(e,e'pn)$, $(\gamma,pp)$, and $(\gamma,pn)$ must be performed.
 The advantage of  pp-emission is the fact that the electromagnetic 
 ``background'', represented  by two-body currents, consists only of the 
 $\Delta$-excitation and deexcitation mechanisms,  whereas the nonrelativistic 
 seagull- and pion-in-flight meson-exchange currents (MEC)
 are forbidden due to isospin selection rules. On the other hand,
 pn-emission, where MEC show rather larger 
contributions\cite{Gui98,Gui99}, allows one to study in addition the tensor 
correlations (TC), due to the strong tensor components of the pion-exchange
contribution of the  NN-interaction.  
Concerning the electromagnetic probe, in electron scattering  a large 
sensitivity to correlations in the longitudinal response has been found 
\cite{Gui97}, whereas in photoabsorption the  only existing transverse part 
is dominated, in most of the kinematics studied till now, by medium-range 
two-body currents \cite{Gui98}. Therefore, electron induced reactions seem
 preferable to explore NN-correlations while photoreactions, besides giving
 complementary information on correlations, are better suited to investigate
 two-body currents, whose good understanding is essential to disentangle and
 investigate short-range effects. Therefore, both types of reactions are
 interesting and worthwhile to be considered. 

A suitable target for this study is the magic spin 0 nucleus $^{16}O$, 
due to the presence of discrete final states in the excitation energy spectrum
of the residual nuclei, $^{14}C$ and $^{14}N$ in pp- and pn-knockout,
respectively, which are well separated in energy and can therefore be separated 
in experiments with good energy resolution. The spin and isospin quantum 
numbers of the 
residual nuclei determine the quantum numbers of the emitted pair inside the 
target. Since different pair wave functions can be differently affected by SRC
and two-body currents, the experimental separation of specific final states can
act as a filter to disentangle and separately investigate the two reaction
processes \cite{Gui97}. This is a peculiar feature of  $^{16}O$
 compared, for example, to a  few-nucleon target like  $^3He$,
 where the residual ``nucleus'' consists only of a nucleon 
 without any excitation spectrum.

The existing  microscopic model calculations (see, e.g.,
\cite{Bof96,Gui98,Gui99,Gui97,Gui97i,Gui00,Gui01,Ryc98a,Ryc97,Ryc98b} 
and references therein) are able to give a reasonable and in some cases even 
fair description of the available data \cite{Kes95,Ond97,Ond98,Ros00,Sta00}.
The results obtained till now have confirmed the validity of the direct knockout
mechanism for low values of the excitation energy of the residual nucleus and
have given clear evidence of SRC in the reaction $^{16}O(e,e' pp)$ for the
transition to the $0^+$ ground state of $^{14}C$ \cite {Sta00}.
Due to the complexity of the subject, this result is a great success of the 
experimental and theoretical efforts. However, some discrepancies have been  
found between theory and data. They may be due to the approximations adopted 
in the models  necessary to reduce the complexity of the calculations.
In order to obtain more insight into the two-nucleon knockout process, the 
models should be improved in the near
 future as much as possible. This is of specific importance for the 
interpretation of  the existing as well as of future data
which are expected from already approved proposals at MAMI in  Mainz 
\cite{Ahr97,Arn98}.

A  crucial assumption adopted in the past was the complete
neglect  of the mutual interaction between
 the  two outgoing nucleons (NN-FSI). Only the major contribution of FSI, due to
 the interaction of each of the two outgoing nucleons with the residual nucleus,
 was taken into account in the different models. The guess was that the effect
 of NN-FSI should not be large, at least in the kinematics usually considered in
 the experiments, where the two nucleons are ejected back to back and thus, for
 instance, in superparallel kinematics, where the two nucleons are parallel and 
 antiparallel to the momentum transfer. 
 The superparallel kinematics is of particular interest for theoretical 
 \cite{Gui91} and 
 experimental \cite{Ros00,Ahr97,Arn98} investigations, 
 because a Rosenbluth L/T-separation becomes   possible
 in this kinematics in order to extract the longitudinal structure function
 that is assumed to be most sensitive to SRC.
 A first  calculation on nuclear matter \cite{Kno00} clearly indicates that 
 NN-FSI can be important also in superparallel kinematics. This result has been
 confirmed by our  recent calculation for the exclusive $^{16}O(e,e' pp)$
 reaction \cite{ScB03}.   Intuitively, this result is not very surprising
 because, in contrast to the N-nucleus case, the N-N cross section does not 
 become small at backward scattering angles.

A consistent treatment of FSI would require a genuine three-body approach for
the interaction of the two nucleons and the residual nucleus, which represents a
challenging task.  A first estimate of the role
of NN-FSI within an approximated but more feasible approach has been done
in \cite{ScB03}, where, however, only  a few results for the $^{16}O(e,e' pp)$
 reaction are presented. The present paper is a continuation of this work  
 \cite{ScB03} within the same approach. More details
 of the theoretical treatment are given and more numerical results are
 discussed.  Our investigation is extended to the other channels besides 
 $(e,e' pp)$, i.e., $(e,e'pn)$, $(\gamma,pp)$, and $(\gamma,pn)$. The different
 effects of NN-FSI on the various  electromagnetic reaction mechanisms are 
 worked out in detail. Work is in progress to tackle the full three-body 
 approach.

The paper is organized as follows.  In section \ref{mod}, the main features
of  our model  for two-nucleon knockout \cite{Gui99,Gui97} are shortly reviewed 
 and the different approaches to FSI are discussed.  
 In section \ref{nnfsi}, a detailed description of the 
 numerical treatment of 
 NN-FSI is given.  The numerical results for different reactions in selected 
 kinematical situations are presented in section \ref{res}. Some conclusions  
 and perspectives of possible improvements  and developments
   are given in section \ref{sum}.

\section{The Model}
\label{mod}
The central quantity for the calculation of the cross section of the reaction 
induced by a real or virtual photon, with momentum $\vec{q}$, where two 
nucleons are emitted from a nucleus, is given by the matrix elements of the 
nuclear charge-current operator between initial and final nuclear
 nuclear many-body   states, i.e.,
\begin{equation}\label{eq1}
J^{\mu}(\vec{q}\,) = \int \bra{\Psi_f} \hat{J}^{\mu}(\vec{r}) \ket{\Psi_i}
e^{i \vec{q} \cdot \vec{r}} {\mathrm d}\vec{r}\,\, .
\end{equation}
Bilinear products of these integrals give the components of the 
hadron tensor, whose suitable combinations give all the observables available 
from the reaction process \cite{Bof96}. 

The model is based on the two assumptions of an exclusive reaction for the
transition to a specific discrete state of the residual nucleus and of 
the direct knockout mechanism \cite{Gui97,Gui97i,Gui91}. Thus, we consider a 
direct one-step process where the electromagnetic probe directly interacts 
with the  pair of nucleons that are emitted and the A-2 $\equiv$ B 
nucleons of the residual nucleus behave  as spectators. Recent experiments 
\cite{Ond97,Ond98,Ros00,Sta00,McG95,Lam96,McG98,Wat00} on reactions induced
by real and virtual photon have confirmed the validity of this mechanism for 
low values of the excitation energy of the residual nucleus. 

As a result of these two assumptions, the integrals (\ref{eq1}) 
can be reduced to a form with three main ingredients: the 
two-nucleon overlap function (TOF) between the ground state of the target 
and the final state of the residual nucleus, the nuclear
 current $\hat{j}^{\mu}$ of the two emitted nucleons,
 and the two-nucleon scattering wave 
 function $\ket{\psi_f}$.

\begin{figure}
\centerline{\includegraphics[width=10cm,angle=0]{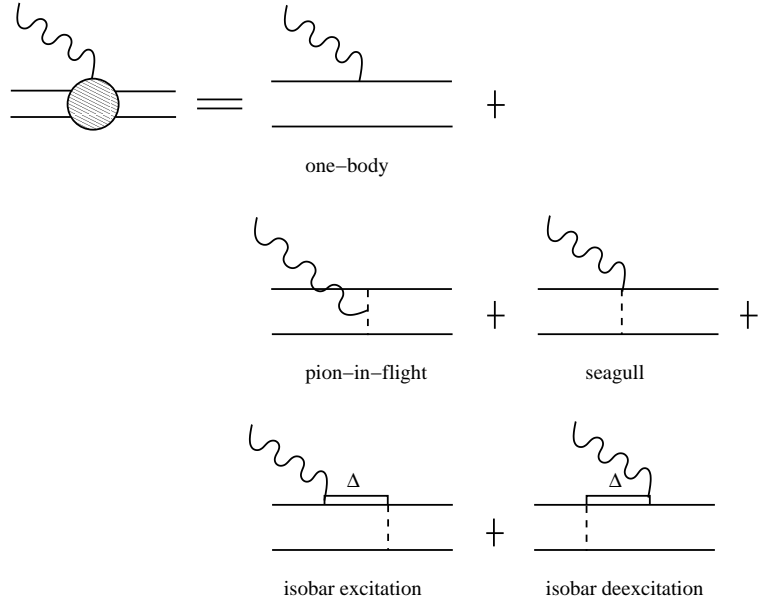}}
\vspace{0.5cm}
\vspace{0.5cm}
\caption{The electromagnetic current contributions taken into account
in the present approach.}
\label{fig_strom}
\end{figure}

The nuclear current operator $\hat{j}^{\mu}(r)$ is the sum of 
 a one- and a two-body part (see fig.\ \ref{fig_strom}). 
 The one-body part consists of  the usual  charge operator and the convection 
and spin currents. In the present version of our model, the two-body part 
consists of the  nonrelativistic  pionic seagull MEC, the pion-in-flight MEC 
and the $\Delta$-contribution, whose explicit expressions can be found, for 
example, in \cite{Gui98,WiA97}. Note that the seagull and the pion-in-flight 
MEC do not contribute in pp-emission, at least in the adopted nonrelativistic 
limit.

The TOF requires a calculation of the two-hole spectral
function including consistently different types of correlations, i.e. SRC
and TC, as well as long-range correlations (LRC), mainly representing
collective excitations of nucleons at the nuclear surface. 

So far, different approaches are available in the most refined version of our
model for pp- and pn-knockout \cite{Gui99,Gui97}. In both cases the TOF for
transitions to discrete low-lying states of  the residual nucleus are given by a
combination of different components of the relative and center-of-mass (CM) 
motion. SRC and TC are introduced in the radial wave function of the relative
motion by means  of state dependent defect functions which are added to the 
uncorrelated partial wave. For the pp-case \cite{Gui97,Geu96}, the defect 
functions are obtained by solving the Bethe-Goldstone equation using, for 
a comparison, different NN-interactions: Bonn OBEPQ-A, Bonn 
OBEPQ-C \cite{Mac89}, and Reid Soft Core \cite{Reid}. 
The calculations with the Bonn OBEPQ-A and Bonn OBEPQ-C potentials do not show 
significant differences, while those with the Reid Soft Core potential 
produce lower cross sections which are in worse agreement with the available 
$(e,e'pp)$ data \cite{Gui97,Ond98,Ros00,Sta00}. For the pn-case \cite{Gui99}, 
SRC and TC correlations are calculated within the framework of the 
coupled-cluster method  \cite{MuP00} with the AV14-potential \cite{WiS84} and 
using the so-called $S_2$ approximation, where only $1$-particle $1$-hole and 
$2$-particle $2$-hole excitations are included in the correlation
operator.  This method is an extension of the Bethe-Goldstone equation and  
takes into account, among other things and  besides particle-particle ladders, 
also hole-hole ladders. These, however, turn out to be rather 
 small  in $^{16}O$ \cite{MuP00}, so that the two approaches are similar in 
 the treatment of SRC. The advantage of the coupled-cluster method is that it
 provides directly correlated two-body wave functions \cite{MuP00,Gui99}.
TC give an important contribution to pn-knockout, while their role is very 
small in pp-knockout.

LRC are included in the expansion coefficients of the TOF. For the pp-case, 
these coefficients are calculated in an extended shell-model basis within a dressed 
random phase approximation \cite{Gui97,Geu96}. For the pn-case, a simple 
configuration mixing calculation of the two-hole
 states in $^{16}O$ has been done and only $1p$-hole states are considered for
 transitions  to the low-lying states of $^{14}N$ \cite{Gui99}.

In  the scattering state, the two outgoing nucleons 1
and 2 and the residual nucleus B interact via the potential
\begin{equation}\label{pot_f}
V_f=V^{OP}(1) + V^{OP}(2) + V^{NN}(1,2),
\end{equation}
where $V^{OP}(i)$ denotes the interaction between the nucleon $i$ and the 
residual nucleus. In our approach we use  a complex phenomenological
 optical potential   fitted to 
  nucleon-nucleus scattering data, which contains
 a central, a Coulomb and a spin-orbit term \cite{Nad81}. 
In order to ensure some consistency in the treatment of the NN-interaction in
the initial and final states, we have used the same NN-potential $V^{NN}$ 
 as in the calculation of the TOF, i.e., the OBEPQ-A potential for pp- and the 
 AV14-potential for pn-emission. The sensitivity of NN-FSI effects to the choice
 of the potential is however small in the calculations.

In general, a consistent treatment of the final state would require a genuine 
three-body approach, that, due to the complexity of the problem, has never
 been  realized till now for complex nuclei.
 Different approximations have been used  in the past.

In the simplest picture, any interaction between the two nucleons and the
residual nucleus is neglected, i.e. $V_f \equiv 0$, and  a plane-wave 
approximation (PW) is assumed for the outgoing nucleon wave functions. 
If $\vec{p}^{\,\,0}_i$, $i \in \{1,2 \}$,  denotes the asymptotic momentum 
of the outgoing nucleon $i$ in the chosen reference frame,
 the corresponding state is therefore given by
\begin{equation}\label{pw1}
 \ket{\psi_f}^{PW} =
 \ket{\vec{p}^{\,\,0}_1}\,\ket{\vec{p}^{\,\,0}_2}\,\, , 
\end{equation}
where $\ket{\vec{p}^{\,\,0}_i}$ describes  the plane-wave state of the nucleon
  $i$ with  momentum $\vec{p}^{\,\,0}_i$.

In a more sophisticated  approach,   only the
 optical potential  $V^{OP}(i)$ is considered
while the mutual NN-interaction $V^{NN}(1,2)$ is neglected.
In this so-called ``distorted wave'' (DW)  approximation, which has 
commonly been used in our previous work, the final state is in
 general given by
\begin{equation}\label{dw1}
\ket{\psi_f}^{DW} = \left(1 + G_0(z) T^{OP}(z) \right) 
 \ket{\vec{p}^{\,\,0}_1}\,\ket{\vec{p}^{\,\,0}_2}\,\, ,
\end{equation}
with
\begin{equation}\label{z}
z= \frac{(\vec{p}^{\,\,0}_1)^2}{2m_1} + \frac{(\vec{p}^{\,\,0}_2)^2}{2m_2}
+ \frac{(\vec{p}_B)^2}{2m_{B}} - i \epsilon \,\, .
\end{equation}
In (\ref{dw1})   $m_i$  denotes the mass of the nucleon $i$ and 
$\vec{p}_B$ is the asymptotic momentum of the residual nucleus, with mass
$m_B$, that in the laboratory frame is given by
\begin{equation}\label{pb}
\vec{q} = \vec{p}^{\,\,0}_1 +  \vec{p}^{\,\,0}_1 +  \vec{p}_B  \,\,.
\end{equation}

The quantity
\begin{equation}\label{g0}
G_0(z) = \frac{1}{z-H_0(1)-H_0(2) -H_0(A-2)}
\end{equation} 
in (\ref{dw1})
denotes the free three-body propagator, where the kinetic Hamiltonian 
$H_0$ is treated nonrelativistically excluding rest masses. 
The scattering amplitude $T^{OP}$ in (\ref{dw1}) is in general given by the 
Lippmann-Schwinger equation, i.e.,
\begin{eqnarray}
T^{OP}(z) &=& \left(V^{OP}(1) + V^{OP}(2) \right)  
 + \left(V^{OP}(1) + V^{OP}(2) \right)  G_0(z) T^{OP}(z) \,\, . \label{top}
\end{eqnarray}

The residual nucleus has a rather large mass in comparison with the
nucleon and can thus be considered as infinitely heavy. This is a good
approximation, which considerably simplifies the calculation of the scattering 
state, 
that can thus be expressed  in this limit as the  product of two uncoupled 
single-particle  distorted wave functions, i.e., 
\begin{equation}\label{dw2}
 \ket{\psi_f}^{DW} =
 \ket{\phi^{OP}(\vec{p}^{\,\,0}_1)}\,\ket{\phi^{OP}(\vec{p}^{\,\,0}_2)},
\end{equation}
where $\ket{\phi^{OP}(\vec{p}^{\,\,0}_i)}$, is given by 
\begin{equation}\label{dw3}
 \ket{\phi^{OP}(\vec{p}^{\,\,0}_i)}  = 
\left(1 + g^i_0(z_i)  t^{OP,i}(z_i) \right)  \ket{\vec{p}^{\,\,0}_i},
\end{equation}
and
\begin{eqnarray}
z_i &=& \frac{(\vec{p}^{\,\,0}_i)^2}{2m_i} - i \epsilon \,\, ,
\label{z1}\\ 
g^i_0(z_i) &=& \frac{1}{z_i-H_0(i)} \,\, , \label{goi}\\
 t^{OP,i}(z_i) &=&  V^{OP}(i) + V^{OP}(i)  g^i_0(z_i) t^{OP,i}(z_i) \,\,.
\label{top_i}
\end{eqnarray}
are the single-particle counterparts to (\ref{z}),
(\ref{g0}),(\ref{top}), respectively.
Equation (\ref{dw3}) is equivalent to the single-particle Schr\"o\-dinger equation
\begin{equation}\label{op-equation}
\left( H_0(i) + V^{OP}(i) \right) \ket{\phi^{OP}(\vec{p}^{\,\,0}_i)} =
\frac{(\vec{p}^{\,\,0}_i)^2}{2m_i}
\ket{\phi^{OP}(\vec{p}^{\,\,0}_i)}
\end{equation}
which is  solved in our treatment numerically in configuration space.

\begin{figure}
\centerline{\includegraphics[width=16cm,angle=0]{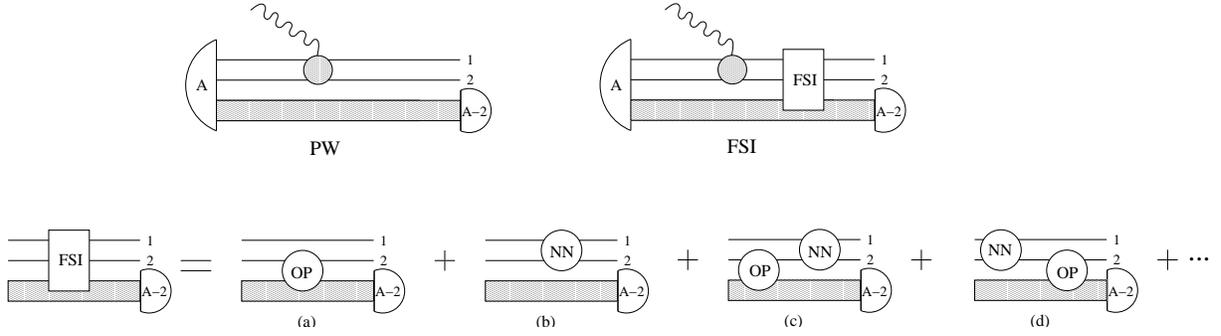}}
\vspace{0.5cm}
\caption{The relevant diagrams for electromagnetic 
 two-nucleon knockout on a complex nucleus
A. The two diagrams on top depict the plane-wave approximation 
 (PW)  and  the distortion of the two outgoing proton wave functions
by final state interactions (FSI).  Below, the relevant mechanisms of 
FSI are depicted in detail, where the open circle denotes either the  
 nucleon-nucleus scattering amplitude (OP), see (\protect{\ref{top_i}}), 
or the
nucleon nucleon-scattering amplitude  (NN), see (\protect{\ref{tnn}}).
Diagrams which are given by an interchange of nucleon 1 and 2 are not depicted.
In the present approach, only diagrams (a) and (b) are taken into account.
}
\label{fig_mechanism}
\end{figure}

In all previous works, but in our recent paper \cite{ScB03},
 the interaction  $V^{NN}$   between   the two outgoing nucleons (NN-FSI) 
 has been completely neglected.
If we want to incorporate it in a fully consistent frame, an infinite series
 of contributions has to be taken into account in $T^{OP}$ and 
 in the NN-scattering amplitude 
\begin{equation}\label{tnn}
t^{NN}(z_{12}) = V^{NN} + V^{NN} g^{12}_0(z_{12}) t^{NN}(z_{12}),
\end{equation}
where $z_{12} = z_1 +  z_2$, and  
\begin{equation}\label{go12}
g^{12}_0(z_{12}) = \frac{1}{z_{12}-H_0(1)-H_0(2)} \,\, . 
\end{equation}
Adopting  again the  
approximation $m_B \rightarrow \infty$, the leading order  terms in the 
scattering amplitudes   are given by (see fig.\ \ref{fig_mechanism})
\begin{eqnarray}
 \ket{\psi_f} &=& \left(1 + g^{12}_0(z_{12}) t^{OP,12}(z_{12})  
+ g^{12}_0(z_{12}) t^{NN}(z_{12}) + 
  g^{12}_0(z_{12}) t^{OP,12}(z_{12})   g^{12}_0(z_{12}) t^{NN}(z_{12})
+ \right. \nonumber \\ & &  \left.  \,\,\,\,\,\,\,\,\,
  g^{12}_0(z_{12}) t^{NN}(z_{12})   g^{12}_0(z_{12}) t^{OP,12}(z_{12})
+ ... \right)  \ket{\vec{p}^{\,\,0}_1}\,\ket{\vec{p}^{\,\,0}_2} \,\, ,
\label{dwnn}
\end{eqnarray}
where $t^{OP,12}(z_{12})$ follows from (\ref{top})  with the 
substitution
\begin{equation}
G_0(z) \rightarrow g^{12}_0(z_{12}) \,\, .
\label{sub1}
\end{equation}

A consistent treatment of FSI would require a
genuine  three-body approach,  with the usual computational challenges,
 by summing up the infinite series (\ref{dwnn}).
 We intend to tackle this project  in the near future.
 At the moment, following the same approach as in \cite{ScB03},
we restrict ourselves to a perturbative treatment by taking into account 
only the first three terms in (\ref{dwnn}),  i.e.\ the plane wave
contribution and diagrams (a) and (b) in  
fig.\ \ref{fig_mechanism}. 
 Formally, this corresponds
to a perturbative treatment of $t^{OP,12}$ and $t^{NN}$ up to first
order and where multiscattering processes, like the fourth and
fifth terms in (\ref{dwnn}) (diagrams (c) and (d) in fig.\
  \ref{fig_mechanism}),
 are neglected. Such an approximated but much more feasible
treatment should allow us to study at least the  main features of NN-FSI. 
In particular, it should allow us to answer the open question whether
 the neglect of NN-FSI in previous calculations can in general be justified
 or  not.  The present treatment of incorporating NN-FSI is denoted as 
 DW-NN. We denote as PW-NN the treatment where only $V^{NN}$ is 
 considered and $V^{OP}$ is switched off.

We would like to add that in 
practice the finite mass $m_B$ of the residual nucleus 
 is taken into 
account  in the PW- and DW-calculations 
 by performing in (\ref{op-equation})
the transformation \cite{Gui91} ($i \neq j$)
\begin{equation}\label{trafo1}
\vec{p}^{\,\,0}_i \rightarrow \vec{q}^{\,\,0}_i = 
\frac{1}{m_{^{16}O}}\left[
(m_j + m_B ) \vec{p}^{\,\,0}_i - m_i (\vec{p}^{\,\,0}_j + \vec{p}_B)
\right],
\end{equation}
 where $m_{^{16}O}$ denotes the mass of the $^{16}O$-target. Moreover, a semirelativistic generalization of
 (\ref{op-equation}) has been used as discussed in
 \cite{Nad81}.
 
In conclusion, the corresponding final states in the
 different approximations are given by
\begin{eqnarray}
 \ket{\psi_f}^{PW} &=&
 \ket{\vec{q}^{\,\,0}_1}\,\ket{\vec{q}^{\,\,0}_2}\,\, , \label{fsi-approx1} \\
 \ket{\psi_f}^{DW} &=&
 \ket{\phi^{OP}(\vec{q}^{\,\,0}_1)}\,\ket{\phi^{OP}(\vec{q}^{\,\,0}_2)}
\,\, , \label{fsi-approx2}\\
\ket{\psi_f}^{PW-NN} &=& \ket{\vec{q}^{\,\,0}_1}\,\ket{\vec{q}^{\,\,0}_2}
+  g^{12}_0(z_{12}) t^{NN}(z_{12}) 
 \ket{\vec{p}^{\,\,0}_1}\,\ket{\vec{p}^{\,\,0}_2}\,\, ,
\label{fsi-approx3} \\
\ket{\psi_f}^{DW-NN} &=& 
 \ket{\phi^{OP}(\vec{q}^{\,\,0}_1)}\,\ket{\phi^{OP}(\vec{q}^{\,\,0}_2)}
+    g^{12}_0(z_{12}) t^{NN}(z_{12}) 
 \ket{\vec{p}^{\,\,0}_1}\,\ket{\vec{p}^{\,\,0}_2}\,\, .
\label{fsi-approx4} 
\end{eqnarray}

\section{Numerical treatment of NN-FSI}
\label{nnfsi}
In general, we intend to be as flexible as possible in the treatment 
of NN-FSI. Consequently,  we use the momentum and not  the configuration space
 for the evaluation of  the scattering amplitude $t^{NN}(z_{12})$  in 
 (\ref{tnn}), so that also nonlocal potentials, like the Bonn OBEPQ-A 
 potential, which is used in \cite{Gui99,Geu96} to calculate the defect 
 functions of the pp-case, can be considered.
 Moreover, in momentum space  it is  much easier to incorporate in the NN-FSI
 also the $\Delta$-isobar consistently within a coupled-channel  
 approach \cite{ScA01}.
 Due to the rather large
 energies of the real or virtual photon in the kinematics discussed below,
 it cannot be excluded  from the beginning 
 that this contribution can be neglected  in the NN-FSI. In forthcoming studies
 we intend to investigate this question in some detail. In the present paper, 
 however, we restrict ourselves to the  genuine NN-potential AV14 
for pn- and the Bonn OBEPQ-A potential for pp-emission.

 Since both the electromagnetic current and the TOF
 are calculated in the model in the configuration space, we have to adopt
 a suitable Fourier transformation of the relative coordinates\cite{Gui91}
 $\vec{r}_{1B}= \vec{r}_{1} - \vec{r}_{B}$ and
 $\vec{r}_{2B}= \vec{r}_{2} - \vec{r}_{B}$ of the three-body system.
Taking into account the CM correction (\ref{trafo1}), we have
 therefore to evaluate
\begin{eqnarray}
\braket{\psi_f}{\vec{r}_{1B}, \vec{r}_{2B}}
 & \equiv&
 \braket{\psi_f(\vec{p}_1^{\,\,0},\vec{p}_2^{\,\,0})}{
\vec{r}_{1B}, \vec{r}_{2B}} = 
\braket{\phi^{OP}(\vec{q}_1^{\,\,0})}{\vec{r}_{1B}}\, \,
\braket{\phi^{OP}(\vec{q}_2^{\,\,0})} {\vec{r}_{2B}}\, \,
+
\nonumber \\
& & \,\,\,   \int d^3p_1 d^3p_2 
\bra{\vec{p}_1^{\,\,0}, \vec{p}_2^{\,\,0}} \,\,
t^{NN}(z_{12}^{\ast}) \,\, g^{12}_0(z_{12}^{\ast}) 
  \ket{\vec{p}_1, \vec{p}_2\,}
\braket{\vec{p}_1}{\vec{r}_{1B}}\,
\braket{\vec{p}_2}{\vec{r}_{2B}}\,\, .
\label{final4_2}
\end{eqnarray}

Note that the final state is completely determined by the asymptotic momenta 
$\vec{p}_1^{\,\,0}$ and $\vec{p}_2^{\,\,0}$ of the two nucleons, while
the recoil momentum  of the residual nucleus $\vec{p}_{\rm B}$ is 
given by Eq. (\ref{pb}). Additional spin and isospin quantum numbers 
are suppressed at the moment for the sake of simplicity.

The 6-dimensional integral in (\ref{final4_2}) can be reduced to a 
3-dimensional one by exploiting in $t^{NN}$ the conservation of the total 
momentum of the two nucleons, i.e.,

\begin{equation}\label{zwischen13}
\bra{\vec{p}_1^{\,\,\prime}, \vec{p}_2^{\,\,\prime}} \,\,
t^{NN}(z_{12}^{\ast})  \ket{\vec{p}_1, \vec{p}_2\,}
= \delta^{(3)}\left(\vec{p}_1^{\,\,\prime} + \vec{p}_2^{\,\,\prime}
  -  \vec{p}_1 -  \vec{p}_2 \right) 
\bra{\frac{ \vec{p}_1^{\,\,\prime} -  \vec{p}_2^{\,\,\prime}}{2}} \,\,
t^{NN}(z_{12}^{\ast})  \ket{\frac{\vec{p}_1 - \vec{p}_2}{2}} \,\, . 
\end{equation}
After some straightforward algebra one obtains 

\begin{eqnarray}
 \braket{\psi_f(\vec{p}_1^{\,\,0},\vec{p}_2^{\,\,0})}{\vec{r}_{1B
}, \vec{r}_{2B}} &=& 
\braket{\phi^{OP}(\vec{q}_1^{\,\,0})}{\vec{r}_{1B}}\, \,
\braket{\phi^{OP}(\vec{q}_2^{\,\,0})} {\vec{r}_{2B}}\, \,
+ \nonumber \\ & &
 \!\!\!\!\!\!\!\!\!\!\!\!\!\!\!\!\!\!\!\!\!\!
\!\!\!\!\!\!\!\!\!\!\!\!\!\!\!\!\!\!\!\!\!\!
 \!\!\!\!\!\!\!\!\!\!\!\!\!\!\!\!\!\!\!\!\!\!
\left(2\pi\right)^{-3} \,\,
e^{-i \frac{\left( \vec{p}_1^{\,\,0} + \vec{p}_2^{\,\,0} \right)}{2} \cdot
       \left(\vec{r}_{1B}+ \vec{r}_{2B} \right) }
\int d^3p
\bra{\vec{p}_{rel}^{\,\,0}} \,\,
 t^{NN}({p}_{rel}^{0}) \ket{\vec{p}\,}
\frac{1}{
\frac{\left({p}_{rel}^{0}\right)^2}{ m_N} -
\frac{p^2}{ m_N} + i \epsilon} 
e^{-i \vec{p}\cdot \left( \vec{r}_{1B} - \vec{r}_{2B} \right)},
\label{final5}
\end{eqnarray}
where the notation $p \equiv \left|\vec{p} \, \right| $ for any vector
$\vec{p}$ is used, and 

\begin{equation}\label{zwischen20}
\vec{p}_{rel}^{\,\,0} = \frac{\vec{p}_1^{\,\,0} - \vec{p}_2^{\,\,0}}{2}
\end{equation}
is the relative momentum  of the two outgoing nucleons.
Due to the fact that the Coulomb force is not incorporated in 
the NN-potential, we use in the propagator of (\ref{final5}) an average value
for the nucleon mass, i.e.. $m_1 = m_2 \equiv m_N =938.926$ MeV.
Moreover, we note that the states in the matrix element 
$t^{NN}(z_{12}^{\ast}) \equiv  t^{NN}({p}_{rel}^{0})$ of (\ref{zwischen13}) and 
(\ref{final5}) correspond only to the relative
motion of the two nucleons. In detail, the scattering amplitude in momentum 
space  is given by the following integral equation
\begin{eqnarray}
\bra{\vec{p}^{\,\,\prime}} \, t^{NN}({p}_{rel}^{0}) \ket{\vec{p}\,}
&=&
\bra{\vec{p}^{\,\,\prime}} \, V_{NN}  \ket{\vec{p}\,} \,+ \,
\int d^3k
\bra{\vec{p}^{\,\,\prime}} \, V_{NN} \ket{\vec{k}\,} 
\frac{1}{
\frac{\left({p}_{rel}^{0}\right)^2}{ M_N} -
\frac{k^2}{ M_N} + i \epsilon} 
\bra{\vec{k}\,} \,  t^{NN}({p}_{rel}^{0}) \ket{\vec{p}\,} \,\, ,
\nonumber \\
\label{tnn_2}
\end{eqnarray}
 which can be solved with standard numerical methods
 \cite{ScZ74,Glo83}. Further details are given below.

In practice, (\ref{final5}) is exploited by using a partial wave decomposition. 
We must consider at this point that also the spin and isospin quantum numbers 
must be given  for the outgoing particles. This means  that 
 the final state   $\ket{\psi_f}$, which we have specified till now only by
\begin{equation}\label{zwischen22}
\ket{\psi_f} \equiv 
 \ket{\psi_f(\vec{p}_1^{\,\,0},\vec{p}_2^{\,\,0})} \,\, ,
\end{equation}
must be extended as
\begin{equation}\label{zwischen23}
\ket{\psi_f} \equiv 
 \ket{\psi_f(\vec{p}_1^{\,\,0},\vec{p}_2^{\,\,0};
  s\, m_s, t \,t_0; \beta)}\,\, .
\end{equation}
Here, $s$ denotes the total spin of the two outgoing  nucleons, with
 projection $m_s$ on their relative momentum 
 $\vec{p}_{rel}^{\,\,0}$ (see (\ref{zwischen20})), $t$
 the total isospin of the two nucleons, with  
third component $t_0$, and $\beta$ includes  
 the  spin and isospin quantum numbers of the residual nucleus state as well as 
 its excitation energy.

 Due to the fact that our chosen reference frame $\Sigma_q$
 has the photon momentum $\vec{q}$ as 
 quantization axis, one has to perform at first an (active) rotation of the
 spin state:
\begin{eqnarray}
\bra{s m_s} &=& \sum_a D^{[s]}_{m_s a}(0, -\Theta, -\Phi) \bra{s a},
\label{zwischen24}
\end{eqnarray}
where  $\Theta$ and $\Phi$ denote the polar angles of
 $\vec{p}_{rel}^{\,\,0}$ in   $\Sigma_q$ and  
the rotation matrices can be found in \cite{Mes61}.
Concerning the optical potential wave functions, 
$\braket{\phi(\vec{q}_1^{\,\,0})}{\vec{r}_{1B}}$ and 
$\braket{\phi(\vec{q}_2^{\,\,0})} {\vec{r}_{2B}}$ in (\ref{final5}),
 an uncoupled basis for the spin of the nucleons is appropriate.
 It is therefore  useful to rewrite  (\ref{zwischen24}) as follows:
\begin{eqnarray}
\bra{s m_s} &=& \sum_{a,s_1,s_2} D^{[s]}_{m_s a}(0, -\Theta, -\Phi) 
\braket{\frac{1}{2}s_1 \frac{1}{2}s_2}{s a}
\bra{\frac{1}{2} s_1}\, \bra{\frac{1}{2} s_2} \,\, .
\label{zwischen25}
\end{eqnarray}
Note that the projection numbers $a, s_1$ and $s_2$ now refer to the
 reference frame $\Sigma_q$, with the photon momentum $\vec{q}$ along
 the $z$-axis, and  $s_1$ ($s_2$) is the spin projection of
 nucleon 1 (2) in  $\Sigma_q$.
Inserting now spin and isospin degrees of freedom into (\ref{final5}),
 we obtain 

\begin{eqnarray}
\braket{\psi_f}{\vec{r}_{1B}, \vec{r}_{2B}}
& \equiv& 
 \braket{\psi_f(\vec{p}_1^{\,\,0},\vec{p}_2^{\,\,0};
  s\, m_s, t \,t_0; \beta)}{\vec{r}_{1B}, \vec{r}_{2B}}
\nonumber \\
& &  
\!\!\!\!\!\!\!\!\!\!\!\!\!\!\!\!\!\!\!\!\!\!
\!\!\!\!\!\!\!\!\!\!\!\!\!\!\!\!\!\!\!\!\!\!
= 
\sum_{a,s_1,s_2} \sum_{t_1,t_2}  \sum_{s'_1,s'_2} \sum_{t'_1,t'_2} 
D^{[s]}_{m_s a}(0, -\Theta, -\Phi) 
\braket{\frac{1}{2}s_1 \frac{1}{2}s_2}{s a} 
\braket{\frac{1}{2}t_1 \frac{1}{2}t_2}{t t_0} \,\,  \times 
 \nonumber \\
 & & 
\!\!\!\!\!\!\!\!\!\!\!\!\!\!\!\!\!\!\!\!\!\!
\!\!\!\!\!\!\!\!\!\!\!\!\!\!\!\!\!\!\!\!\!\!
\braket{\phi(\vec{q}_1^{\,\,0},s_1 s'_1,t_1 t'_1; \beta)}{\vec{r}_{1B}}\, \,
\braket{\phi(\vec{q}_2^{\,\,0},s_2 s'_2,t_2 t'_2; \beta)} {\vec{r}_{2B}}\, \,
\,\, \bra{\frac{1}{2} s'_1}\,\, \bra{\frac{1}{2} s'_2}\,\,
\bra{\frac{1}{2} t'_1}\,\, \bra{\frac{1}{2} t'_2}
+ \nonumber \\ & &
\!\!\!\!\!\!\!\!\!\!\!\!\!\!\!\!\!\!\!\!\!\!
\!\!\!\!\!\!\!\!\!\!\!\!\!\!\!\!\!\!\!\!\!\!
\left(2\pi\right)^{-3} \,\,
e^{-i \frac{\left( \vec{p}_1^{\,\,0} + \vec{p}_2^{\,\,0} \right)}{2} \cdot
       \left(\vec{r}_{1B}+ \vec{r}_{2B} \right) }
\int d^3p
\bra{\vec{p}_{rel}^{\,\,0},s \, m_s, t\, t_0} \,\,
 T_{NN}({p}_{rel}^{0}) \ket{\vec{p}\, }
\frac{1}{
\frac{\left({p}_{rel}^{0}\right)^2}{ M_N} -
\frac{p^2}{ M_N} + i \epsilon} 
e^{-i \vec{p}\cdot \left( \vec{r}_{1b} - \vec{r}_{2B} \right)}
\,\, , \nonumber \\
\label{final6}
\end{eqnarray}
where  the distorted wave function
$\braket{\phi(\vec{q}_i^{\,\,0},s_i,s'_i,t_i,t'_i; \beta)}{\vec{r}_{i3}}$
with spin and isospin projection numbers $s_i,s'_i,t_i,t'_i$ is given by
 (consider (\ref{dw3}))

\begin{equation}\label{top3}
\braket{\phi(\vec{q}_i^{\,\,0},s_i,s'_i,t_i,t'_i; \beta)}{\vec{r}_{i3}}
= \bra{\vec{q}_i^{\,\,0},s_i,t_i; \beta}
\left(1+ t^{OP,i}(z^{\ast}_i) g_0^i(z^{\ast}_i) \right)
\ket{\vec{r}_{i3},s'_i,t'_i}\,\, .
\end{equation}

In order to exploit (\ref{final6}), we use the following 
identities:
\begin{itemize}
\item partial wave decomposition of a plane wave:
\end{itemize}
\begin{equation}\label{zwischen30}
e^{i \vec{a}\cdot \vec{b}}= 4\pi \sum_{l,m} i^l j_l(ab) 
\left(Y^{[l]}_m(\hat{a})\right)^{\ast}  Y^{[l]}_m(\hat{b}), 
\end{equation}
where $j_l$ denote the spherical Bessel functions;

\begin{itemize}
\item  partial wave decomposition of $\bra{\vec{p}_{rel}^{\,\,0},s \, m_s}$
 in (\ref{final6}):

\begin{equation}\label{zwischen31}
 \bra{\vec{p}_{rel}^{\,\,0}, s \, m_s}= 
 \frac{1}{\sqrt{4\pi}} \sum_{ljm} \sqrt{2l+1} \braket{l0sm_s}{jm_s}
 D^{[j]}_{m_sm}(0,-\Theta,-\Phi)
 \bra{p^0_{rel} (ls)jm}\,\, ,
\end{equation}
where  the projection $m$ refers to the $z$-axis of $\Sigma_q$;

\end{itemize}
\begin{itemize}
\item  partial wave decomposition of a state $\ket{\vec{p}\,}$

\begin{eqnarray}
\braket{\vec{p}\,} { (q(ls)jm,t t_0)}&=&
 \frac{1}{p^2}\delta(p-q)
\left[ Y^{[l]}(\hat{p}) \times \left[ \chi^{[\frac{1}{2}]}(1)
 \times \chi^{[\frac{1}{2}]}(2)
\right]^{[s]} \right]^{[j]}_m \,\, \ket{\frac{1}{2} \frac{1}{2} t t_0}
\,\, ,  \label{zwischen34}
\end{eqnarray}
 where $\chi^{[\frac{1}{2}]}(i)$ denotes the Pauli-spinor of nucleon $i$.
\end{itemize}

\begin{itemize}
\item 
For the matrix element of the $NN$-scattering amplitude
 between partial waves, one can exploit the fact that $t^{NN}$ is a rank-0 
 tensor operator:
\begin{equation}\label{zwischen26}
\bra{p^{\prime} (l^{\prime}s^{\prime}) j^{\prime} m^{\prime} ,
 t^{\prime} t_0^{\prime}}  t^{NN}(p^0_{rel}) \ket{p (ls) j m ,t t_0}
= \delta_{j j^{\prime}}  \delta_{m m^{\prime}} \delta_{t t^{\prime}}
 \delta_{t_0 t_0^{\prime}} \delta_{s^{\prime} s}  
t^{j\, s\, t}(p^{\prime} l^{\prime},p l; p^0_{rel})  \,\, .
\end{equation}
The quantity $t^{j\, s\, t}(p^{\prime} l^{\prime},p l; p^0_{rel})$
is given by a 1-dimensional integral equation which can be derived
from (\ref{tnn_2}). With the help of Gaussian mesh points,
 the latter can be transformed into a matrix equation which is solved 
 directly by matrix inversion \cite{Sch99}.
\end{itemize}

With the help of these relations, one obtains, after some straightforward
 algebra, the following final result:

\begin{eqnarray}
\braket{\psi_f}{\vec{r}_{1B}, \vec{r}_{2B}}
& \equiv& 
 \braket{\psi_f(\vec{p}_1^{\,\,0},\vec{p}_2^{\,\,0};
  s\, m_s, t \,t_0; \beta)}{\vec{r}_{1B}, \vec{r}_{2B}}
\nonumber \\
& &  
\!\!\!\!\!\!\!\!\!\!\!\!\!\!\!\!\!\!\!\!\!\!
\!\!\!\!\!\!\!\!\!\!\!\!\!\!\!\!\!\!\!\!\!\!
=  \,\,
\sum_{a,s_1,s_2} \sum_{t_1,t_2}  \sum_{s'_1,s'_2} \sum_{t'_1,t'_2} 
D^{[s]}_{m_s a}(0, -\Theta, -\Phi) 
\braket{\frac{1}{2}s_1 \frac{1}{2}s_2}{s a} 
\braket{\frac{1}{2}t_1 \frac{1}{2}t_2}{t t_0} \,\,  \times 
 \nonumber \\
 & & 
\!\!\!\!\!\!\!\!\!\!\!\!\!\!\!\!\!\!\!\!\!\!
\!\!\!\!\!\!\!\!\!\!\!\!\!\!\!\!\!\!\!\!\!\!
\braket{\phi(\vec{q}_1^{\,\,0},s_1 s'_1,t_1 t'_1; \beta)}{\vec{r}_{1B}}\, \,
\braket{\phi(\vec{q}_2^{\,\,0},s_2 s'_2,t_2 t'_2; \beta)} {\vec{r}_{2B}}\, \,
\,\, \bra{\frac{1}{2} s'_1}\,\, \bra{\frac{1}{2} s'_2}\,\,
\bra{\frac{1}{2} t'_1}\,\, \bra{\frac{1}{2} t'_2}
+ \nonumber \\ & &
\!\!\!\!\!\!\!\!\!\!\!\!\!\!\!\!\!\!\!\!\!\!
\!\!\!\!\!\!\!\!\!\!\!\!\!\!\!\!\!\!\!\!\!\!
\left(2\pi\right)^{-3} \,\,
e^{-i \frac{\left( \vec{p}_1^{\,\,0} + \vec{p}_2^{\,\,0} \right)}{2} \cdot
       \left(\vec{r}_{1B}+ \vec{r}_{2B} \right) }
\sum_{s'_1,s'_2} \sum_{t'_1,t'_2}
\sum_{l^{\prime},n_1,n_2} \sqrt{4\pi} F_{l^{\prime},n_1,n_2}(\left|
\vec{r}_{1B} - \vec{r}_{2B} \right|, p^0_{rel}, s m_s,t)  \,\, \times 
 \nonumber \\
 & & 
\!\!\!\!\!\!\!\!\!\!\!\!\!\!\!\!\!\!\!\!\!\!
\!\!\!\!\!\!\!\!\!\!\!\!\!\!\!\!\!\!\!\!\!\! 
Y^{[l^{\prime}]}_{-n_1}(\widehat{\vec{r}_{2B}-\vec{r}_{1B}})
\,\,
 \braket{\frac{1}{2} s'_1 \frac{1}{2} s'_2}{s n_2}
 \braket{\frac{1}{2} t'_1 \frac{1}{2} t'_2}{t t_0}
\,\,\bra{\frac{1}{2} s'_1}\,\, \bra{\frac{1}{2} s'_2}\,\,
\bra{\frac{1}{2} t'_1}\,\, \bra{\frac{1}{2} t'_2} 
\label{final7}
\end{eqnarray} 
 where the function $F$ in (\ref{final7}) is given by
\begin{eqnarray}
 F_{l^{\prime},n_1,n_2}(\left|\vec{r}_{1B} - \vec{r}_{2B} \right|,
 p^0_{rel}, s m_s,t) &=& 
\sum_{l,j} \int dp p^2 i^{l^{\prime}}
j_{l^{\prime}}(p \left|\vec{r}_{2B}-\vec{r}_{1B}\right|)
(-1)^{n_1}
\frac{1}{
\frac{\left({p}_{rel}^{0}\right)^2}{ M_N} -\frac{p^2}{ M_N} + i \epsilon} 
\nonumber \\ & & 
\!\!\!\!\!\!\!\!\!\!\!\!\!\!\!\!\!\!\!\!\!\!\!\!\!\!\!\!\!\!\!\!\!\!\!\!\!\!\!
\!\!\!\!\!\!\!\!\!\!\!\!\!\!\!\!\!\!\!\!\!\!\!\!\!\!\!\!\!\!\!\!\!\!\!\!\!\!\!
\sqrt{2l+1} \braket{l0sm_s}{jm_s}  D^{[j]}_{m_s n_1 + n_2}(0,-\Theta,-\Phi)
t^{j\, s\, t_0}(p^{\prime} l^{\prime},p l; p^0_{rel}) 
\braket{l^{\prime} n_1 s n_2}{j n_1+n_2} \,\, . 
\label{final8}
\end{eqnarray}

 In our explicit evaluation,  we have taken an upper limit of
 $l^{\prime} =3$,
 i.e., we have considered the isospin-1 partial waves 
$^1\!S_0, ^3\!P_0, ^3\!P_1, ^3\!P_2, ^1\!D_2, ^3\!F_2, ^3\!F_3$,
 and $^3\!F_4$ for pp-knockout and in addition the isospin-0 contributions
 $^3\!S_1, ^1\!P_1,  ^3\!D_1, ^3\!D_2,  ^3\!D_3$, and $^1\!F_3$ 
for pn-knockout. 
 Concerning the summation index $l$ in  (\ref{final8}), 
 the limit  $l^{\prime}  \leq 3$ implies that in addition also 
 $l=4$ and $l=5$ contributions ($^3\!G_3$ and $^3\!H_4$) are considered.
 It has been checked numerically that this truncation
 is sufficient at least   for the kinematics considered
 in this paper.

\section{Results}
\label{res}
In this section, we discuss the role of NN-FSI on different electromagnetic
reactions with pp- and pn-knockout from $^{16}O$. The case of the
$^{16}O(e,e'pp)^{14}C$ reaction has already been considered in \cite{ScB03}, 
where it has been found that the effects of NN-FSI depend on kinematics,
on the different partial waves for the relative motion of the nucleon pair in 
the initial state and, therefore, on the final state of the 
residual nucleus. In particular, a considerable enhancement for medium and large
values of the recoil momentum  has been found, for the transition to the $0^+$ 
ground state of $^{14}C$, just in the superparallel kinematics of a recent 
experiment at MAMI \cite{Ros00}. Since similar kinematics has been proposed
for the first $^{16}O(e,e'pn)^{14}N$ experiment at MAMI \cite{Arn98}, this is
the first case we have considered in the present investigation. 

\begin{figure}
\begin{tabular}{cc}
\centerline{\includegraphics[width=16cm,angle=0]{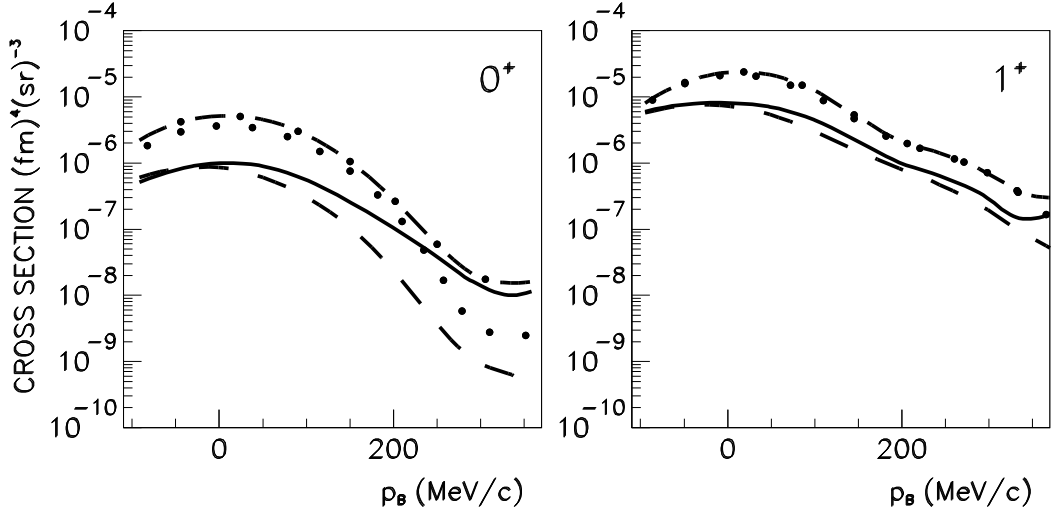}}

\end{tabular}
\vspace{0.5cm}

\caption{The differential cross section of the $^{16}O(e,e' pp)$ reaction to 
the $0^+$ ground state of $^{14}C$ (left panel) and of the $^{16}O(e,e' pn)$ 
reaction to the $1^+$ ground state of $^{14}N$ (right panel) in a  
superparallel kinematics  with an incident electron energy $E_0= 855$ MeV,  
an electron scattering angle $\theta_e = 18^{\circ}$, energy transfer 
$\omega=215$ MeV and  $q=316$ MeV/$c$.
In $^{16}O(e,e' pn)$ the proton is ejected parallel and the neutron 
antiparallel to $\vec{q}$.  
 Different values of $p_{\rm B}$ are obtained changing the kinetic energies of the outgoing
nucleons. Positive (negative) values of $p_{\rm B}$ refer to situations where 
${\vec p}_{\rm B}$ is parallel (anti-parallel) to ${\vec q}$.
 Line convention:
 PW (dotted), PW-NN (dash-dotted), DW (dashed), DW-NN (solid).
}
\label{result1}
\end{figure}

The calculated  differential cross sections of the $^{16}O(e,e'pp)$ reaction to
 the  $0^+$ ground state of $^{14}C$ and of the $^{16}O(e,e'pn)$ reaction to 
 the $1^+$ ground state of $^{14}N$ in superparallel kinematics 
are displayed in the left and right panels
 of fig.\ \ref{result1}, respectively.  The results given by the different 
 approximations 
 (\ref{fsi-approx1}-\ref{fsi-approx4}) are
compared in the figure. The $^{16}O(e,e'pp)^{14}C$ cross section was already 
presented in \cite{ScB03} and is shown again here only to allow a more direct
comparison of FSI effects on pp- and pn-emission in the same kinematics. 

It can be clearly seen in the figure that the inclusion of the optical potential
leads, in both reactions, to an overall and substantial reduction of the
calculated cross sections (see the difference between the PW and DW results).
This effect is well known and it is mainly due to the imaginary part of 
the optical potential, that accounts for the flux lost to inelastic channels in
the nucleon-residual nucleus elastic scattering. The optical potential gives the
dominant contribution of FSI for recoil-momentum values up to  
$p_{\rm B} \simeq 150$ MeV/$c$. At larger values NN-FSI gives an enhancement of 
the cross section, that increases with $p_{\rm B}$. In $(e,e'pp)$ this 
enhancement 
goes beyond the PW result and amounts to roughly an order of magnitude for 
$p_{\rm B} \simeq 300$ MeV/$c$. In $(e,e'pn)$ this effect is still sizeable
 but  much weaker.  
We note that in both cases the contribution of NN-FSI is larger in the DW-NN 
than in the PW-NN approximation.

\begin{figure}
\centerline{\includegraphics[width=10cm,angle=0]{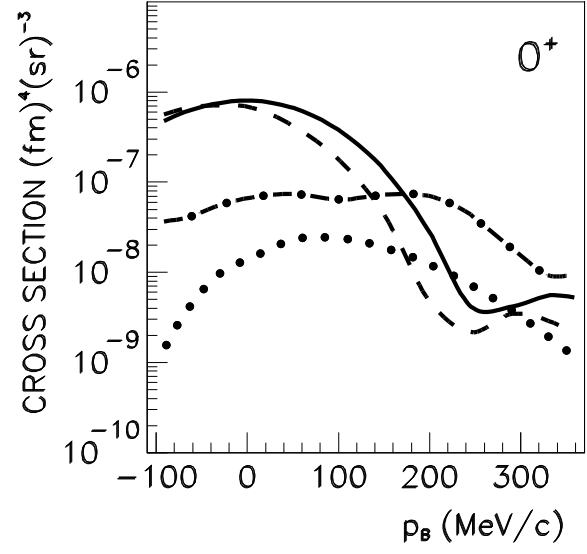}}
\vspace{0.5cm}
\caption{The differential cross section of the $^{16}O(e,e' pp)$ reaction to 
the $0^+$ ground state of $^{14}C$
 in the same superparallel kinematics as in fig.\ \protect{\ref{result1}}.
Line convention:
 DW with the $\Delta$-current (dotted), 
 DW-NN with the $\Delta$-current (dash-dotted),
 DW with the one-body current (dashed),
 DW-NN with the one-body current (solid).
}
\label{result2}
\end{figure}

\begin{figure}
\centerline{\includegraphics[width=16cm,angle=0]{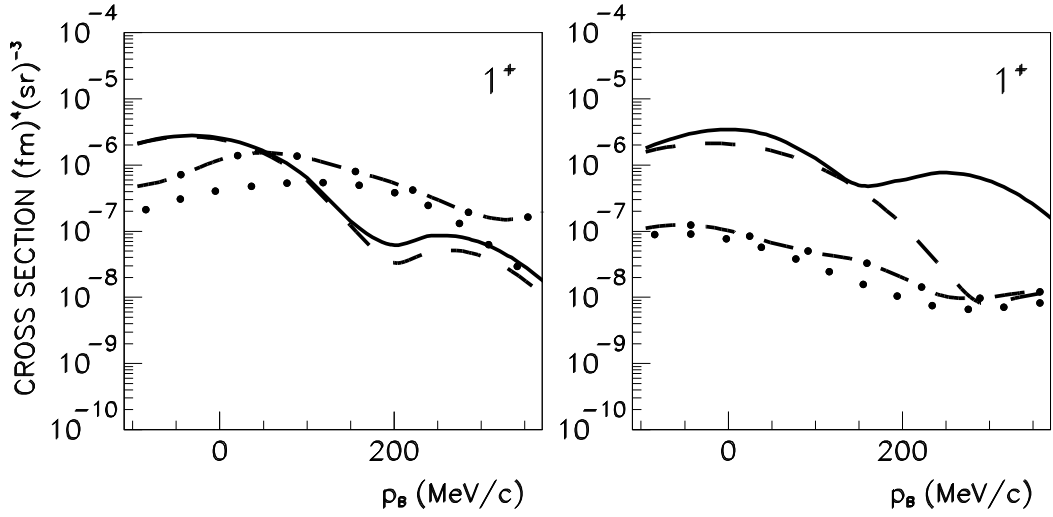}}
\vspace{0.5cm}
\caption{The differential cross section of the $^{16}O(e,e' pn)$ reaction to 
the $1^+$ ground state of $^{14}N$ 
 in the same superparallel kinematics as in fig.\ \protect{\ref{result1}}.
 Line convention in the left panel:
 DW with the $\Delta$-current (dotted), 
  DW-NN with  the $\Delta$-current (dash-dotted),
 DW  with the one-body-part (dashed),
 DW-NN  with the one-body-part (solid).
 Line convention in the right panel:
 DW  with the pion-in-flight-current (dotted), 
 DW-NN with the pion-in-flight--current (dash-dotted),
 DW with the seagull-current (dashed),
 DW-NN with  the seagull-current (solid).
}
\label{result3}
\end{figure}

In order to understand NN-FSI effects  in some more detail,
the separated contributions of the different terms of the nuclear current 
in the DW and DW-NN approximations are compared  in fig.\ \ref{result2} for 
pp- and in fig.\  \ref{result3} for pn-knockout.
In $(e,e'pp)$  NN-FSI produces a strong enhancement of the $\Delta$-current 
contribution for all the values of $p_{\rm B}$. Up to about 100-150 MeV/$c$, 
however, this effect is completely overwhelmed by the dominant contribution 
of the one-body current, while for larger values of $p_{\rm B}$, where the one-body 
current is less important in the cross section, the increase of the 
$\Delta$-current is responsible for the substantial enhancement in the final 
result of fig.\ \ref{result2}. 
The effect of NN-FSI on the one-body current is much weaker but anyhow 
 sizeable, and it is responsible for the NN-FSI effect at lower values of 
$p_{\rm B}$ in fig.\ \ref{result2}.

We have found in a detailed analytical analysis that the antisymmetrization of 
the final NN-state leads to a strong suppression of the $\Delta$-current 
contribution if  the momenta of  the nucleons after the deexcitation of the 
$\Delta$  (consider fig.\  \ref{fig_strom})  
 are aligned  parallel or antiparallel to the photon momentum $\vec{q}$. 
 However, this situation is in general not present when NN-FSI are
  taken into account, even in superparallel kinematics. 

Different effects of NN-FSI on the various components of the current are
shown for the $(e,e'pn)$ reaction in fig.\ \ref{result3}. Also in this
case, NN-FSI affects more the two-body than the one-body current. A sizeable
enhancement is produced on the $\Delta$-current, at all the values of 
$p_{\rm B}$, 
and a huge enhancement on the seagull current at large momenta. In contrast, 
the one-body current is practically unaffected by NN-FSI up to about 150 
MeV/$c$. A not very large but visible enhancement is produced at larger 
momenta, where, however, the one-body current gives only a negligible
contribution to the final cross section. The role of the pion-in-flight 
term, in both DW and DW-NN approaches, is practically negligible in the cross
section. Thus, a large effect is given by NN-FSI on the seagull and the 
$\Delta$-current. The sum of the 
two terms, however, produces a destructive interference that leads to a partial 
cancellation in the final cross section. The net effect of NN-FSI in 
fig.\ \ref{result1} is not large but anyhow non negligible. Moreover, the
results for the partial contributions in fig.\ \ref{result3} indicate that
in pn-knockout NN-FSI can be large in particular situations and therefore 
should in general be included in a careful evaluation. 

\begin{figure}
\centerline{\includegraphics[width=8cm,angle=0]{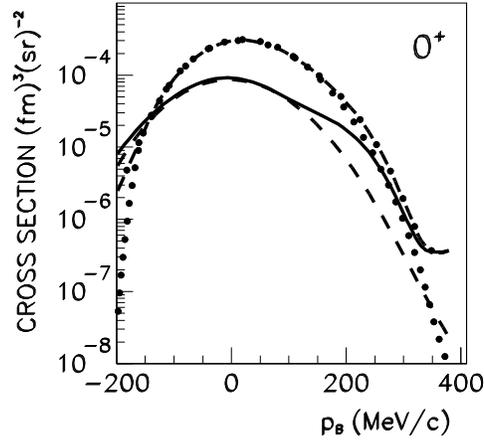}}
\vspace{0.5cm}
\caption{The differential cross section of the $^{16}O(\gamma,pp)$ reaction to 
the $0^+$ ground state of $^{14}C$ in superparallel kinematics at 
$E_\gamma =$ 215 MeV. 
 Line convention  as in fig.\  \protect{\ref{result1}}.
}
\label{result4}
\end{figure}

\begin{figure}
\centerline{\includegraphics[width=8cm,angle=0]{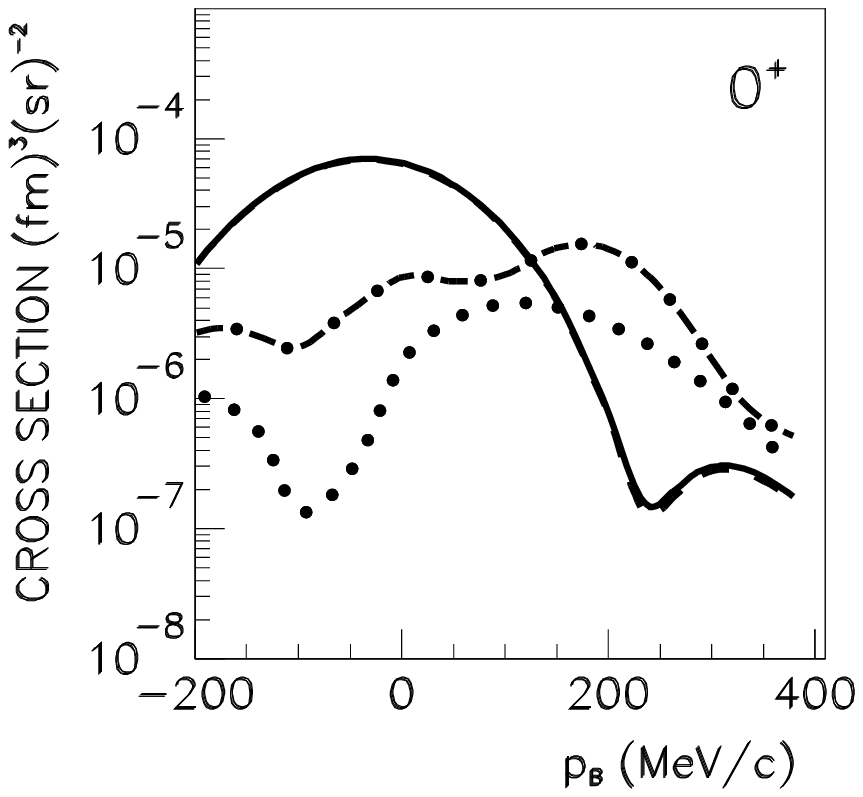}}
\vspace{0.5cm}
\caption{The differential cross section of the $^{16}O(\gamma,pp)$ reaction to 
the $0^+$ ground state of $^{14}C$ in the same kinematics as in 
fig.\ \protect{\ref{result4}}. 
 Line convention  as in 
 fig.\  \protect{\ref{result2}}. 
}
\label{result5}
\end{figure}

The cross section of the $^{16}O(\gamma,pp)$ reaction to the  $0^+$ ground 
state of $^{14}C$ calculated with the different approximations for FSI 
is shown in fig.\ \ref{result4}. The separated contributions of the one-body 
and $\Delta$-currents in DW and DW-NN  are displayed in fig.\ \ref{result5}. 
Calculations have been performed in superparallel kinematics, and for an
incident photon  energy which has the same value, $E_\gamma= 215$  MeV, as 
the energy transfer in the $(e,e'pp)$ calculation of fig.\ \ref{result1}. 
This kinematics, which is not very well 
suited for $(\gamma,pp)$ experiments, can be interesting for a
theoretical comparison with the corresponding results of the electron induced
reactions in figs.\ \ref{result1} and \ref{result2}.

In general, two-body currents give the major contribution to $(\gamma,NN)$ 
reactions. In this superparallel kinematics, however, the  $(\gamma,pp)$ 
cross section is dominated by the one-body current for recoil momentum values 
up to about 150 MeV/$c$. For larger values the $\Delta$-current plays the main
role. This is the same behavior as in the corresponding situation for 
$(e,e'pp)$. In fig.\ \ref{result5} NN-FSI produces a significant 
enhancement of the $\Delta$-current contribution, that, however, is not 
as large as in $(e,e'pp)$. The role 
of NN-FSI on the one-body current is negligible in $(\gamma,pp)$, while it is 
significant in $(e,e'pp)$. This effect is produced in $(e,e'pp)$ on
the longitudinal part of the nuclear current, that does not contribute in reactions
induced by a real photon. 
Thus, in practice, in this kinematics NN-FSI affects only the $\Delta$-current 
and therefore in fig.\ \ref{result4} its effect is negligible in the region 
where the one-body current is dominant. At large values of $p_{\rm B}$, where 
the role of the $\Delta$-current becomes important, the enhancement produced 
by NN-FSI is sizeable, but weaker
than in the same superparallel kinematics for $(e,e'pp)$.
We note that also for the $(\gamma,pp)$ reaction in fig.\ \ref{result4} NN-FSI 
effects are larger in the DW-NN than in the PW-NN approach.

\begin{figure}
\begin{tabular}{cc}

\end{tabular}
\vspace{0.5cm}
\centerline{\includegraphics[width=14cm,angle=0]{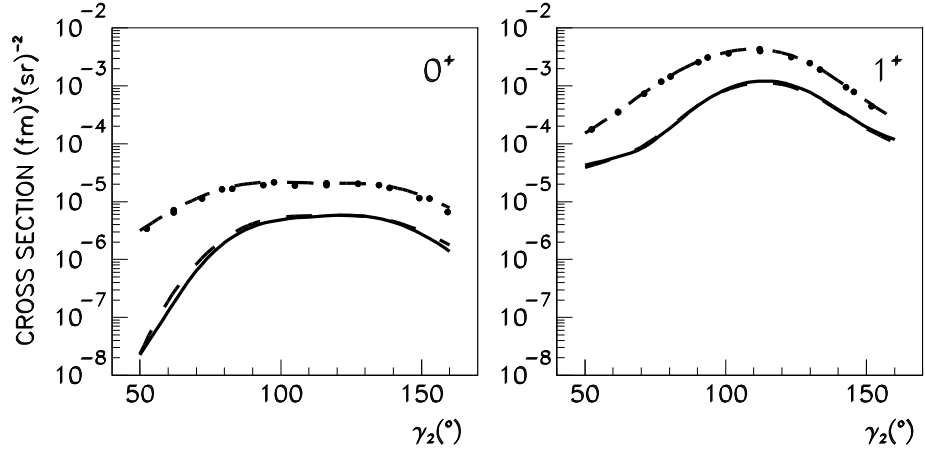}}
\caption{The differential cross section  of 
  the $^{16}O(\gamma, pp)$ reaction to the $0^+$ ground state of $^{14}C$ 
  (left panel) and of the $^{16}O(\gamma, pn)$ reaction to the $1^+$ ground 
  state of $^{14}N$ (right panel)  as a function of the scattering angle 
  $\gamma_2$ of the second outgoing nucleon in a coplanar kinematics with 
  $E_\gamma =$ 120 MeV, $T_1 = 45$ MeV and $\gamma_1=45^{\circ}$. 
  Line convention as in fig.\  \protect{\ref{result1}}.
 }
\label{result6}
\end{figure}

\begin{figure}
\centerline{\includegraphics[width=8cm,angle=0]{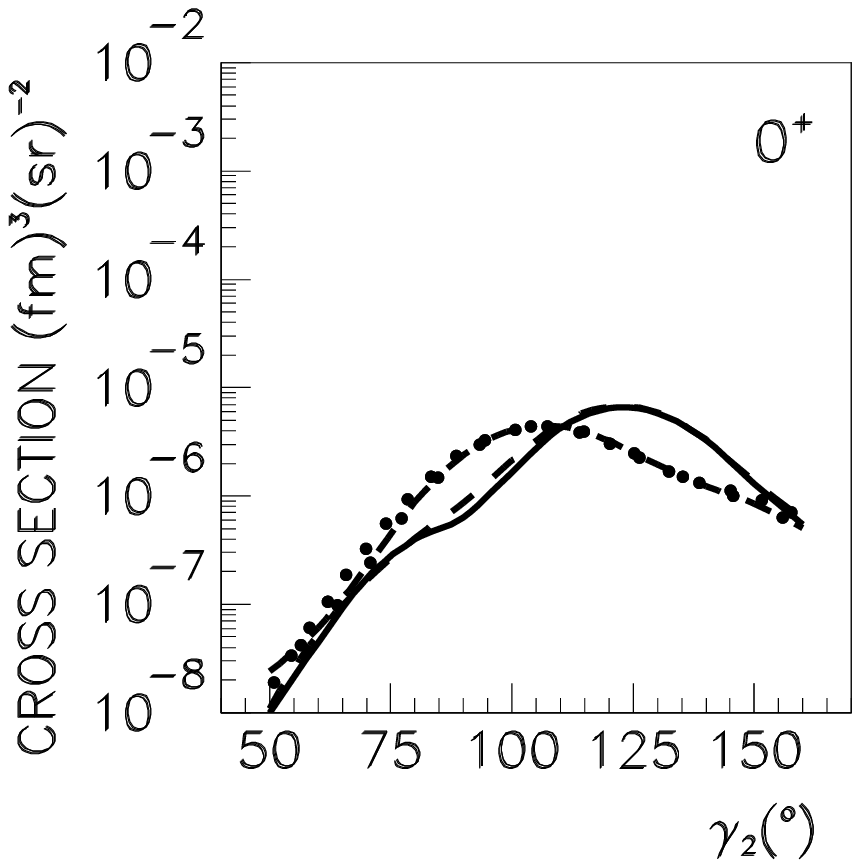}}
\vspace{0.5cm}
\caption{The differential cross section  of 
  the $^{16}O(\gamma, pp)$ reaction to the $0^+$ ground state of $^{14}C$ 
  in the same kinematics as in fig.\ \protect{\ref{result6}}.
  Line convention as in fig.\ \protect{\ref{result2}}.
  }
\label{result7}
\end{figure}

Another example is presented in fig.\ \ref{result6}, where the results of the
different approximations in the treatment of FSI are displayed for the  
$^{16}O(\gamma,pp)$ reaction to the  $0^+$ ground state of $^{14}C$ (left panel)
and for the $^{16}O(\gamma,pn)$ reaction to  the $1^+$ ground state of $^{14}N$ 
(right panel) in a coplanar kinematics at $E_\gamma = 120$ MeV, where the 
energy and the scattering angle of the outgoing proton are fixed at 
$T_1= 45$ MeV and $\gamma_1= 45^{\circ}$, respectively. Different values of 
the recoil momentum  can be obtained by varying the scattering angle $\gamma_2$ 
of the second outgoing nucleon on the other side of the
 photon momentum.  It can be clearly seen in the figure 
that NN-FSI has almost no effect. In contrast, a very large contribution is 
given, for both reactions, by the optical potential, which produces a 
substantial reduction of the calculated cross sections. 
This kinematics, which appears within reach of available experimental
facilities,  was already envisaged in \cite{Gui01} as promising to study SRC in
the $(\gamma,pp)$ reaction. In fact, at the considered value of the photon
energy, the contribution of the $\Delta$-current is relatively much less 
important, and while the $(\gamma,pn)$ cross section is dominated by the 
seagull current \cite{Gui01}, in the $(\gamma,pp)$ cross section the 
contribution of the one-body current is large and competitive with the one of 
the two-body current. This can be seen in fig.\ \ref{result7}, 
where the two separated contributions are shown in the DW and  in the DW-NN
approximations. Both processes are important: the $\Delta$-current plays the
main role at lower values of $\gamma_2$, while for  $\gamma_2 \geq 110^{\circ}$
the one-body current and therefore SRC give the major contribution. The effect
of NN-FSI is practically negligible on both terms, which explains the result in
the final cross section of fig.\ \ref{result7}.

A study of the $(\gamma,pp)$ reaction in a kinematics of the type considered
 in figs.\ \ref{result6} and \ref{result7}, where NN-FSI is negligible and
 correlations are important, might represent a promising alternative to the 
 $(e,e'pp)$ reaction for the investigation of SRC.

\section{Summary and Outlook}
\label{sum}

The relevance of the mutual final state interaction of the two emitted nucleons 
(NN-FSI) in electro- and  photoinduced two-nucleon knockout from  $^{16}O$ has
been investigated 
 within a perturbative treatment. 

A consistent evaluation of FSI would require a genuine three-body approach, for
the two nucleons and the residual nucleus, by summing up an infinite series of
contributions  in the NN-scattering amplitude and in the interaction of the two
nucleons with the residual nucleus. In all previous calculations performed till
now on complex nuclei,
 apart from our recent paper \cite{ScB03}, NN-FSI was neglected and only 
the interaction of each of the two nucleons
with the residual nucleus was included. In our model this effect is accounted
for by a phenomenological optical potential. Following the same approach as in 
\cite{ScB03}, here NN-FSI has been incorporated in the model within a 
perturbative treatment of the optical potential and of the NN-interaction up to 
first order in the corresponding scattering amplitudes. Therefore, both effects 
of FSI, due to NN-FSI and to the optical potential, are taken into account 
in the present treatment, but the
multiscattering processes, where the two effects are intertwined, are
neglected.  In spite of that, the
most important part of both contributions is presumably included in
the present treatment. Such an approximated but more feasible approach should
therefore be able to give a reliable idea of the relevance of NN-FSI. 

The full three-body approach, which represents a computationally challenging
task, will anyhow be tackled in forthcoming studies, in order to give a more
definite answer about the role of NN-FSI, especially in those situations 
where the effect turns out to be large in the present treatment.

Numerical results of cross sections  calculated for different
reactions and kinematics have been presented.
 In order to understand in more detail the role of the
various effects of FSI, the results of the perturbative treatment, where both
the optical potential and the NN-interaction are included, have been compared
with the more approximated approaches where only either contribution is
considered, as well as with the simplest calculations where FSI are completely
neglected and the PW approximation is assumed for the outgoing nucleons. 

In general, the optical potential gives an overall and substantial reduction of
the calculated cross sections. This important effect represents the main
contribution of FSI and can never be neglected. In  most of 
 the situations
considered here, NN-FSI gives an enhancement of the cross section. The effect 
is in general non negligible, it depends strongly on the kinematics, on the 
type of
reaction, and, as it is shown in \cite{ScB03}, on the final state of the
residual nucleus. NN-FSI affects in a different way the various terms of the
nuclear current, usually more the two-body than the one-body terms, and is
sensitive to the various theoretical ingredients of the calculation. This makes
it difficult to make predictions about the role of NN-FSI in a particular
situation. In general each specific situation should be individually
investigated. 

The results obtained in the present investigation indicate that NN-FSI effects
are in general larger in pp- than in pn-knockout and in electro- than in
photoinduced reactions. In particular situations they can be negligible, 
e.g., in the $^{16}O(\gamma,pp)^{14}C$ and $^{16}O(\gamma,pn)^{14}N$ reactions 
for the coplanar kinematics at $E_\gamma = 120$ MeV
considered here. In particular situations they can be 
large, e.g. in the superparallel kinematics of the $^{16}O(e,e'pp)^{14}C_{\rm
g.s.}$ reaction, where NN-FSI leads to a strong enhancement of the cross
section, up to about one order of magnitude at large values of the recoil
momentum. A qualitatively similar but quantitatively much weaker effect is
obtained, in the same kinematics, for the $^{16}O(e,e'pn)^{14}N$ and 
$^{16}O(\gamma,pp)^{14}C$ reactions. 

In general, NN-FSI is non negligible. In spite of that, the original guess, 
that justified its neglect in the past, i.e. that its contribution does not
significantly change the main qualitative features of the theoretical results,
is basically correct. But if we want to obtain more reliable quantitative
results and to get more insight into the two-nucleon knockout process, for a
more careful comparison with available as well as future data, NN-FSI must be
included in the model. 

In order to improve the reliability of the theoretical description of the
two-nucleon knockout process, the full three-body problem of the final state has
to be tackled in forthcoming studies. In that context, special emphasis has to
be devoted to a more consistent treatment of the initial and the final state.

\vspace{0.2cm} 
\centerline{{\bf Acknowledgements}}
\vspace{0.2cm}
This work has partly been performed under the contract  HPRN-CT-2000-00130 of
the European Commission. Moreover, it has been supported  by 
 the Istituto Na\-zionale di Fisica Nucleare and by the
 Deutsche For\-schungs\-gemein\-schaft (SFB 443). Fruitful discussions with
 H.\ Arenh\"ovel are gratefully acknowledged.

\newpage

\end{document}